\documentclass[modern]{aastex7}

\usepackage{natbib}
\usepackage{comment}
\usepackage{todonotes}
\usepackage{graphicx}
\usepackage{lineno}
\usepackage{comment}
\usepackage{amssymb}
\usepackage{rotating}
\usepackage{amsmath}

\newcommand\lamlam{\mbox{$\:\lambda\lambda $ }}
\newcommand\ha{{H$\alpha$}}

\newcommand\kms{\:\rm{\,km\,s^{-1}}}

\newcommand\hi{\ion{H}{1}}

\newcommand\sii{[\ion{S}{2}]}

\newcommand\oii{[\ion{O}{2}]}
\newcommand\oiii{[\ion{O}{3}]}

\begin{document}

\title{Three-Dimensional Kinematics of the Oxygen-rich Supernova Remnant G292.0+1.8}

\author[0000-0002-9912-5705]{Adele L. Plunkett}
\affil{Department of Physics, Middlebury College, Middlebury, VT 05753, USA}
\affil{National Radio Astronomy Observatory (NRAO), Charlottesville, VA, }
\email{aplunket@nrao.edu}

\author[0000-0001-6311-277X]{ P. Frank Winkler}
\affil{Department of Physics, Middlebury College, Middlebury, VT 05753, USA}
\email{winkler@middlebury.edu}

\author[0000-0002-4134-864X]{Knox S. Long}
\affil{Space Telescope Science Institute, 3700 San Martin Drive, Baltimore, MD 21218, USA; long@stsci.edu}
\affil{Eureka Scientific, Inc.
2452 Delmer Street, Suite 100, Oakland, CA 94602-3017}
\email{long@stsci.edu}

\author[0000-0002-0763-3885]{Dan Milisavljevic}
\affil{Department of Physics and Astronomy, Purdue University, West Lafayette, IN, 47907-2036, USA}
\email {dmilisav@purdue.edu}
\begin{abstract}

Studying the remnants of young core-collapse supernovae (SNe) \added{can yield insight} into the chemical composition of their progenitors and the geometry  of the explosions.  The supernova remnant (SNR) G292.0+1.8 is one of only three known oxygen-rich SNRs in the galaxy--remnants of core-collapse for which relatively pure fragments of ejecta can be seen.  Several dozen ejecta knots from G292.0+1.8 were  the subject of a proper motion analysis, based on  \oiii \ 5007 \AA\  images taken over a 22-year baseline by \citet{winkler09}. They determined that the transverse velocities of the filaments are linearly proportional to their distances from a common expansion center, thus the O-rich filaments have been traveling with little deceleration since the initial supernova event, $\sim 3000$ years ago.   In this paper, we use optical spectra of G292.0+1.8, all taken from the Cerro Tololo Inter-American Observatory (CTIO), to measure radial velocities for 93 knots. Assuming un-decelerated expansion, as indicated by the proper motions, the radial velocity should be proportional to the distance from the center along the line of sight, just as the  proper motions are proportional to the transverse distance.  Therefore, we can map the three-dimensional structure and kinematics of the SNR.  We find that the knots  generally follow  a broad bi-conical  distribution, suggesting that the supernova explosion  produced broad jets of ejecta.  This structure is similar to that seen in some other young core-collapse supernova remnants.

\end{abstract}


\section{Introduction\label{intro}}A young supernova remnant (SNR), resulting from the core collapse and supernova explosion of a massive star, provides insight into the composition of its progenitor star, as well as the kinematics of its explosion mechanism.  
In these young remnants, it is typical to find  fragments of ejecta,  launched from the core of the progenitor star in the explosion, that   appear optically as  knots that have remained virtually uncontaminated through interaction with interstellar or circumstellar material. These knots become visible as they encounter a reverse shock in the evolving SNR, and produce optical spectra dominated by heavy elements, with little or no hydrogen, and velocities that are typically $\gtrsim 1000 \kms$, tracing back to the explosion.   The strongest of the emission lines are usually those of oxygen---the most abundant element in the outer core of typical progenitors.  Remnants containing such knots are commonly known as  O-rich SNRs,  and  include Cas A \citep{kirshner77,chevalier78a}, Puppis A \citep{winkler85}, and G292.0+1.8 \citep{goss79, murdin79} in the Galaxy and a handful of others in the Magellanic Clouds and other nearby galaxies.  Of these, G292.0+1.8 (hereafter G292) is unique in that it simultaneously displays all the characteristics in the canonical picture of a core-collapse SNR: O-rich optical emission from ejecta fragments, apparent circumstellar interaction in a shell produced by an expanding shock, a central neutron star manifested as a pulsar, and a synchrotron-emitting pulsar wind nebula (PWN) powered by an active pulsar \citep{hughes01,camilo02}.  

G292 was first discovered in a radio survey by \citet{mills61}, who denoted it MSH11-5\textit{4}, and its non-thermal spectrum was used to identify G292 as a SNR by \citet{milne69} and \cite{shaver70}.  X-rays from G292 were first identified in HEAO-1 data by \citet{share78}, and further investigations from the \textit{Einstein} Observatory by \citet{clark80} and \citet{tuohy82}, showed a prominent central O-rich ring,  suggesting a massive rotating progenitor.  \citet{park02,park07} used the {\em Chandra} X-Ray Observatory to produce high-resolution ACIS images that revealed the complex morphology of G292, including the central bar with roughly  solar-type composition, the pulsar and its associated  PWN, and the periphery of metal-rich ejecta knots \added{which extend to the outer blast wave.}

\citet{goss79} discovered optical emission from G292, and \citet{murdin79} suggested that the detected ejecta are un-decelerated and not yet contaminated by interstellar material.  These studies first identified G292 as an oxygen-rich SNR.  \citet{gaensler03} used detailed radio images to find that G292 has a diameter of $\sim 8$\arcmin \ and is at a distance $6.2\pm0.9$\ kpc, based primarily on the \hi\ absorption profile.  \citet{ghavamian05} performed an extensive kinematic study using the Rutgers Fabry-Perot imaging spectrometer  to detect the \oiii \ $\lambda$\ 5007 emission features from G292 throughout a velocity range of -1440 $\lesssim v_{rad} \lesssim$1700 km s$^{-1}$.  They found an asymmetric distribution with mostly blue-shifted knots in the north and both red- and blue-shifted knots in the south, as well as a bright, mostly red-shifted eastern spur.  From a plot of radial velocity versus projected radius, they proposed that the knots are distributed around a shell expanding with ejecta shell velocity $v_{ej}\approx1700$\ km s$^{-1}$.  

    The kinematic  study by \citet{ghavamian05} covered the central $7\arcmin$  of  G292, including most of its ejecta knots, and \citet{winkler06} then showed that additional ejecta knots extend throughout most of the 8\arcmin \ shell and its extension to the south,  as seen in radio and X-ray images.  Also, several knots showed \sii $\lambda\lambda$\ 6716, 6731 emission in addition to the oxygen lines, which are evidence of O-burning products in the ejecta.  \citet[][hereafter WTRL]{winkler09}  showed optical images of filaments emitting \oiii \ distributed throughout much of the shell.  With no evidence of hydrogen, these filaments all appear to be fragments of pure SN ejecta.  

Using emission-line images from seven epochs, 1986 to 2008, WTRL found that the trajectories of the ejecta knots are  consistent with a free-expansion model, and clearly show systematic motions
outward from a point near the center of the radio/X-ray shell.  They found a kinematic age of $2990\pm60$\ years, in agreement with the estimated age found by \citet[][3000-3400 yr]{ghavamian05} and also consistent with the pulsar spin-down age \citep[2900 yr,][]{camilo02} and the estimate based on the PWN properties.  

Here we describe our analysis of moderate resolution spectra of knots of \oiii\ emission in G292+1.8, obtained in order to better characterize the  three-dimensional (3-d) structure of the ejecta from the SN explosion, which in turn provides important clues about asymmetries in the SN explosion itself.  The remainder of this paper is organized as follows: In Section 2, we describe the observations and our measurement of the line-of-sight velocities of the knots.  In Section 3, we  convert the transverse velocities obtained by WTRL and our velocity measurements to a 3-d structure, which appears to be dominated by material expanding along a roughly N-S axis. Finally, in Section 4, we summarize our conclusions.  

\section{OBSERVATIONS AND DATA REDUCTION}
\begin{figure}
\begin{center}
\includegraphics[width=6 in]{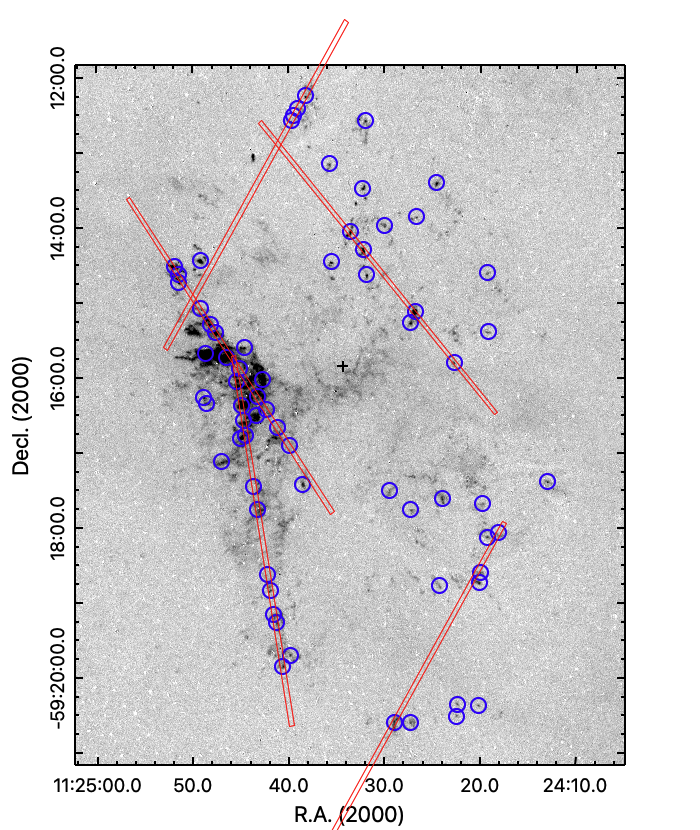}
\end{center}
\caption{Continuum-subtracted \oiii\ image of G292.0+1.8, with the five  R-C spectrograph slit positions  indicated in red.  The 68 spatially distinct knots used in this analysis indicated by blue circles.   These include both knots  lying along the R-C spectrograph slits and those targeted with Hydra. }
\label{slits}
\end{figure}

Spectra used in this study were taken during two separate observing runs at Cerro Tololo Inter-American Observatory (CTIO) in Chile, as described in Table \ref{tab_obs}.  We performed long-slit spectroscopy on 26-28 March 2006 (UT) from the 1.5-m \textit{f}/7.5 telescope, using the Ritchey-Chr\'{e}tien (R-C) spectrograph  and the Loral 1k \#1 CCD with 15 $\mu$ pixels in a 1200 $\times$ 800 array.  The slit was 3\farcs5 wide, with spatial dimension 1\farcs3/pixel. We targeted five slit positions, with four to twelve  knots distinguishable in each position, resulting in a total of 36 knots identified from these observations.  The slit positions are shown in red in Figure  \ref{slits}.

\begin{deluxetable}{lcccccccc}
\tabletypesize{\small}
\tablewidth{0pt}
\tablecaption{Journal of CTIO Spectroscopic Observations\label{tab_obs}}
\tablehead{
\colhead {} &
\colhead {Telescope,} & 
\colhead {} & 
\colhead{Slit/} &
\colhead{} &
\colhead{Wavelength} & 
\colhead{Dispersion} & 
\colhead{} & 
\colhead {}\\ 
\colhead{Date (U.T.)} & 
\colhead {Spectrograph} &
\colhead {CCD} &
\colhead {Field} & 
\colhead{Grating} &
\colhead{Range (\AA)} &
\colhead{${\rm(\AA\; pixel^{-1})}$} &
\colhead {Exposure (s)} &
\colhead {Observers}
 }

\startdata
2006 Mar 27 & 1.5m,  R-C & Loral 1K & N & 32 & 4590--8025 & 2.89 & $4 \times 2500$ & PFW, KSL, \\
	&	&	&	S  &  		&	&	&   $3 \times 2500$ &  K. Twelker \\
	&	&	&	E  &  		&	&	&   $3 \times 1800$ &   \\
\\
2006 Mar 28 &	&	&   N	&  09  &  3540--6970  &  2.88  & $4 \times 3000$  &  \\
	&	&	&	S  &		&	&	&   $2 \times 3000$  &  \\
	&	&	&	NW  &	&	&	&	$2 \times 2500$  &  \\
\\
2006 Mar 29 &	&	&   SE	&  16  &  4841--6849  &  1.69  & $1 \times 1200$  &  \\
	&	&	&	  &		&	&	&    $2 \times 2000$  &  \\
\\	
2008 Mar 30 & 4m, Hydra & SITe 2K$\times$4K & 1 & KPGL3 & 4591--7420 & 1.38 & $3 \times 3000$ & PFW,  ALP \\
2008 Mar 31 & 	 & 	 & 2 & 	& 	 & 	 & $4 \times 3000$  \\
2008 Apr 1 & 	 & 	 & 3 & 	& 	 & 	 & $4 \times 3000$   \\
\enddata

\end{deluxetable}

\begin{figure}
\epsscale{0.85}
\plotone{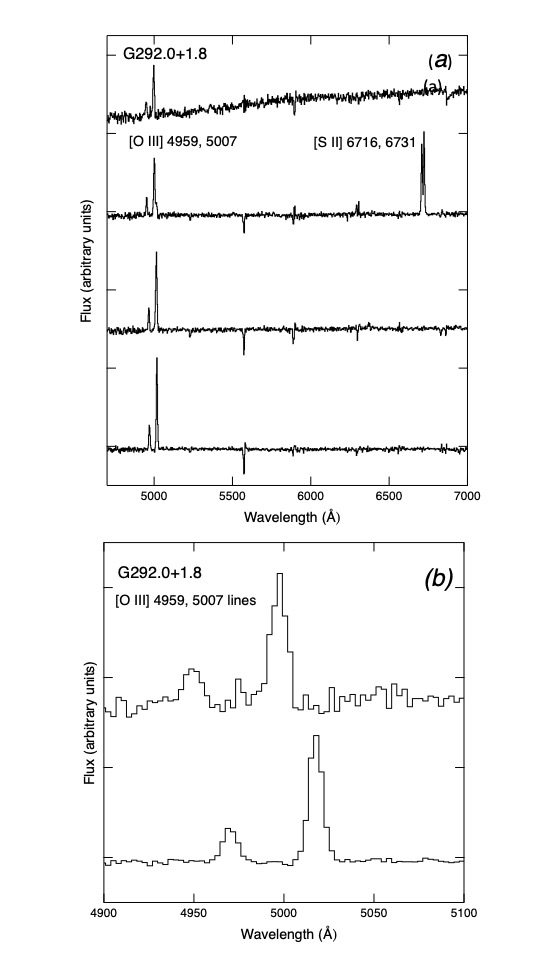}
\caption{(a, top) One-dimensional spectra for four knots located along the northernmost slit.  These spectra were extracted from the same 2-D spectrum, which was the combination of five exposures taken on the first night of observations with the R-C spectrograph on the 1.5m telescope.  These are typical of the G292 knots.  \oiii\ \lamlam 4959, 5007  is prominent in all the knots, while a fraction of them also have strong \sii\ \lamlam 6716, 6731.  Note the complete absence of Balmer lines.  (b, bottom) Enlargement of first and third (from bottom) spectra above, emphasizing the Doppler shifts.The lower spectrum is redshifted, while the upper one is blueshifted.
\label{fig:spectra}}
\end{figure}

In addition to object frames, we took HeAr comparison spectra  for each setup, typically both before and after we observed a particular slit position.  Bias frames, sky flats, and dome flats were taken at the beginning or end of each night of observations.     On each of the three nights we used a different grating, with different spectral coverage, but all included the \oiii \ 
$\lambda\lambda$ \ 4959,5007 lines.  Figure \ref{fig:spectra} shows four sample spectra along the North slit, with clear detection of the \oiii\ lines. The observational set-up details are given in Table \ref{tab_obs}.

On 29-31 March 2008 (UT) we used the Hydra multi-object spectrograph on the 4-m Blanco telescope.  The Hydra ``Bench Schmidt'' camera has 400 mm focal length and uses a SITe 2048 $\times$ 4096 CCD chip with 15 $\mu$ pixels.   We used the KPGL\#3 grating for a spectral coverage of about 2800 \AA; the grating specifications are given in Table \ref{tab_obs}.  Fibers were 300 $\mu$ in diameter, and approximately 130 were functional.  We took several bias frames at the beginning or end of each observing night, along with dome flats  with the  fibers in a circular orientation.   For each field we took  projector flats and HeNeAr comparison spectra were taken for each field, along with ``milky flats''  to remove pixel-to-pixel differences on the chip.

During three nights, we observed three fields of 19 to 21 distinct knots each and several more sky targets, described in Table \ref{tab_obs} and shown in Figure \ref{slits}.  We observed some objects  in more than one Hydra field and some that we had previously observed  along one of the R-C slit positions.  Upon reduction, we found that several of the knots comprised two or more kinematic components.  A total of 93 kinematically distinct knots were observed and spectroscopically identified.  


We reduced the R-C data using standard IRAF\footnote{NOIRLab IRAF is distributed by the Community Science and Data Center at NSF NOIRLab, which is managed by the Association of Universities for Research in Astronomy (AURA) under a cooperative agreement with the U.S. National Science Foundation.} procedures for bias-subtraction and flat-fielding.  For Hydra spectra, we used the IRAF reduction task {\tt dohydra}, which allows for scattered light subtraction, extraction, flat fielding, fiber throughput correction, wavelength calibration, and sky subtraction in a single complete data reduction path.

We extracted one-dimensional spectra  for all identified knots using the task {\tt blkavg}.  At least the two strong \oiii\  4959 and 5007 \AA\ emission-lines were consistently apparent in all the 1D spectra, and the \sii \ 6716 and 6731 \AA\ emission-lines were apparent for several knots.  As is typical for O-rich remnants, Balmer lines of H are absent or nearly so in all the knots.  For consistency, we measured only the \oiii \ doublet for the kinematic analysis.  For each spectrum, we fit the \oiii \ emission-line features with two Gaussians of equal width in order to measure the centroid wavelengths and Gaussian full-widths at half-maximum (FWHM).  We estimate that we can determine the centroid velocity of each feature with an accuracy of 1/20 of the line width, so the velocities that we measured have an accuracy of $\pm$ FWHM/20 in velocity units.

In cases where what we had designated as a single knot based on the continuum-subtracted image had two or three velocity components in its spectrum,  we fit all features with Gaussians, and we considered the pairs of features with wavelength difference of $\sim 48$ \AA\ (the difference in rest wavelength of the \oiii \ doublet) to stem from the same knot.   We designate such kinematically distinct knots by a, b, c, ... in Table~\ref{tab:vrad}.

\clearpage

\startlongtable
\begin{deluxetable*}{lcccrrrrrrrrl}
\tablewidth{0pt}

\tabletypesize{\scriptsize}
\tablecaption{Knot positions and radial velocities
\label{tab:vrad}}
\tablehead{ 
\colhead{Label\tablenotemark{a}}	&	\colhead{Field\tablenotemark{b}}	&	\colhead{R.A.}		&	\colhead{Decl.}	&	\colhead{$\Delta x$\tablenotemark{c}}	&	\colhead{$\Delta y$\tablenotemark{c}}	&	\colhead{$v_z$}	&	\colhead{$\Delta x$\tablenotemark{d}}	&	\colhead{$\Delta y$\tablenotemark{d}}	&	\colhead{$\Delta z$\tablenotemark{e}}	&	\colhead{R\tablenotemark{f}}	&	\colhead{V\tablenotemark{g}}	& \colhead{Lines\tablenotemark{h}} \\
\colhead{}	&	\colhead{}	&	\colhead{(J2000.)}	&	\colhead{(J2000.)}	&		\colhead{($\arcsec$)}	&	\colhead{($\arcsec$)}	&	\colhead{($\kms$)}	&	\colhead{(pc)}	&	\colhead{(pc)}	&	\colhead{(pc)}	&	\colhead{(pc)}	&	\colhead{(km s$^{-1}$)} & \colhead{}
}
\startdata
1	&	H	&	11\phn 24\phn 	12.9	&--59\phn17\phn23.2	&--164.9	&--92.2	&	219.1	&--4.8	&--2.7	&	0.7	&	5.5	&	1805.6 & O, S	   \\
2	&	RC	&	11\phn 24\phn 	18.0	&--59\phn18\phn	04.5	&--125.7	&--133.5	&	838.3	&--3.7	&--3.9	&	2.6	&	5.9	&	1930.7	& O\\
3	&	H	&	11\phn 24\phn 	19.1	&--59\phn18\phn 07.9& --117.1	&--136.9	&	954.6	&--3.4	&--4.0	&	2.9	&	6.0	&	1957.3	& O\\
4	&	H	&	11\phn 24\phn 	19.1	&--59\phn15\phn23.7	&--116.9	&	27.3	&--705.2	&--3.4	&	0.8	&--2.2	&	4.1	&	1340.0& O, S	\\
5	&	H	&	11\phn 24\phn 	19.3	&--59\phn14\phn36.5	&--116.1	&	74.5	&--394.8	&--3.4	&	2.2	&--1.2	&	4.2	&	1367.3	& O\\
6	&	H	&	11\phn 24\phn 	19.7	&--59\phn17\phn41.3	&--112.4	&--110.3	&	487.4	&--3.3	&--3.2	&	1.5	&	4.8	&	1571.6	& O\\
7	&	H, RC	&	11\phn 24\phn 	19.9	&--59\phn18\phn36.5	&--110.9	&--165.5	&	1199.1	&--3.2	&--4.8	&	3.7	&	6.9	&	2238.0	& O, S \\
8	&	H, RC	&	11\phn 24\phn 	20.0	&--59\phn18\phn44.5	&--110.1	&--173.5	&	1192.4	&--3.2	&--5.0	&	3.7	&	7.0	&	2285.0	& O, S\\
9	&	H	&	11\phn 24\phn 	20.1	&--59\phn20\phn22.7	&--109.4	&--271.7	&	347.5	&--3.2	&--7.9	&	1.1	&	8.6	&	2800.1	& O, S\\
10	&	H	&	11\phn 24\phn 	22.3	&--59\phn20\phn21.5	&--92.5	&--270.5	&	283.1	&--2.7	&--7.9	&	0.9	&	8.4	&	2726.8	& O\\
11	&	H	&	11\phn 24\phn 	22.4	&--59\phn20\phn31.3	&--91.9	&--280.3	&	355.5	&--2.7	&--8.2	&	1.1	&	8.6	&	2821.0	& O, S\\
12	&	H	&	11\phn 24\phn 	22.7	&--59\phn15\phn48.1	&--90.0	&	2.9	&--1034.5	&--2.6	&	0.1	&--3.2	&	4.1	&	1341.5	& O\\
13	&	H	&	11\phn 24\phn 	23.9	&--59\phn17\phn36.9	&--80.4	&--105.9	&--233.3	&--2.3	&--3.1	&--0.7	&	3.9	&	1282.9	& O, S\\
14	&	H	&	11\phn 24\phn 	24.2	&--59\phn18\phn46.9	&--78.2	&--175.9	&	1571.5	&--2.3	&--5.1	&	4.8	&	7.4	&	2409.4	& O\\
15	&	H	&	11\phn 24\phn 	24.6	&--59\phn13\phn24.5	&--75.5	&	146.5	&--82.7	&--2.2	&	4.3	&--0.3	&	4.8	&	1565.9	& O, S\\
16	&	H	&	11\phn 24\phn 	26.6	&--59\phn13\phn51.7	&--59.5	&	119.3	&--98.5	&--1.7	&	3.5	&--0.3	&	3.9	&	1268.8	& O, S\\
17	&	H	&	11\phn 24\phn 	26.7	&--59\phn15\phn07.3	&--58.8	&	43.7	&--1357.4	&--1.7	&	1.3	&--4.2	&	4.7	&	1524.9	& O, S\\
18a	&	H	&	11\phn 24\phn 	27.2	&--59\phn17\phn46.1	&--55.4	&--115.1	&--709.6	&--1.6	&--3.3	&--2.2	&	4.3	&	1404.3	& O, S\\
18b	&	H	&	11\phn 24\phn 	27.2	&--59\phn17\phn46.1	&--55.4	&--115.1	&	51.9	&--1.6	&--3.3	&	0.2	&	3.7	&	1213.0	& O\\
19	&	H	&	11\phn 24\phn 	27.2	&--59\phn20\phn36.3	&--55.2	&--285.3	&	163.8	&--1.6	&--8.3	&	0.5	&	8.5	&	2761.9	& O, S\\
20	&	H	&	11\phn 24\phn 	27.3	&--59\phn15\phn16.3	&--54.6	&	34.7	&--1170.4	&--1.6	&	1.0	&--3.6	&	4.1	&	1321.4	& O\\
21	&	H, RC	&	11\phn 24\phn 	28.9	&--59\phn20\phn36.1	&--42.2	&--285.1	&	41.3	&--1.2	&--8.3	&	0.1	&	8.4	&	2734.7	& O\\
22a	&	H	&	11\phn 24\phn 	29.4	&--59\phn17\phn30.7	&--38.4	&--99.7	&--464.5	&--1.1	&--2.9	&--1.4	&	3.4	&	1114.9	& O\\
22b	&	H	&	11\phn 24\phn 	29.4	&--59\phn17\phn30.7	&--38.4	&--99.7	&--473.9	&--1.1	&--2.9	&--1.5	&	3.4	&	1118.9	& O, S\\
23	&	H	&	11\phn 24\phn 	29.9	&--59\phn13\phn58.3	&--34.2	&	112.7	&--536.1	&--1.0	&	3.3	&--1.6	&	3.8	&	1239.2	& O\\
24a	&	H	&	11\phn 24\phn 	31.8	&--59\phn14\phn37.7	&--20.0	&	73.3	&--704.9	&--0.6	&	2.1	&--2.2	&	3.1	&	1008.2	& O\\
24b	&	H	&	11\phn 24\phn 	31.8	&--59\phn14\phn37.7	&--20.0	&	73.3	&--1081.2	&--0.6	&	2.1	&--3.3	&	4.0	&	1299.4	& O\\
25	&	H	&	11\phn 24\phn 	31.9	&--59\phn12\phn34.3	&--19.0	&	196.7	&	190.9	&--0.6	&	5.7	&	0.6	&	5.8	&	1884.5	& O\\
26	&	H, RC	&	11\phn 24\phn 	32.2	&--59\phn14\phn17.7	&--17.0	&	93.3	&--1006.3	&--0.5	&	2.7	&--3.1	&	4.1	&	1349.8	& O\\
27	&	H	&	11\phn 24\phn 	32.2	&--59\phn13\phn28.7	&--16.8	&	142.3	&--592.9	&--0.5	&	4.1	&--1.8	&	4.5	&	1483.1	& O, S\\
28	&	H, RC	&	11\phn 24\phn 	33.5	&--59\phn14\phn03.1	&--7.0	&	107.9	&--906.6	&--0.2	&	3.1	&--2.8	&	4.2	&	1369.1	& O\\
29	&	H	&	11\phn 24\phn 	35.5	&--59\phn14\phn27.7	&	8.2	&	83.3	&--1045.9	&	0.2	&	2.4	&--3.2	&	4.0	&	1313.2	& O\\
30	&	H	&	11\phn 24\phn 	35.7	&--59\phn13\phn09.3	&	10.2	&	161.7	&--692.2	&	0.3	&	4.7	&--2.1	&	5.2	&	1685.9	& O, S\\
31	&	H, RC	&	11\phn 24\phn 	38.2	&--59\phn12\phn14.9	&	29.1	&	216.1	&	549.6	&	0.8	&	6.3	&	1.7	&	6.6	&	2140.6	& O\\
32	&	H	&	11\phn 24\phn 	38.5	&--59\phn17\phn25.9	&	31.3	&--94.9	&--1267.3	&	0.9	&--2.8	&--3.9	&	4.9	&	1582.6	& O, S\\
33	&	RC	&	11\phn 24\phn 	39.0	&--59\phn12\phn24.7	&	35.0	&	206.3	&	347.7	&	1.0	&	6.0	&	1.1	&	6.2	&	2015.1	& O\\
34	&	RC	&	11\phn 24\phn 	39.4	&--59\phn12\phn31.0	&	38.5	&	200.0	&	314.6	&	1.1	&	5.8	&	1.0	&	6.0	&	1957.8	& O\\
35	&	H, RC	&	11\phn 24\phn 	39.7	&--59\phn12\phn35.1	&	40.5	&	195.9	&--462.8	&	1.2	&	5.7	&--1.4	&	6.0	&	1953.8	& O, S\\
36a	&	H	&	11\phn 24\phn 	39.8	&--59\phn19\phn42.5	&	41.4	&--231.5	&--673.2	&	1.2	&--6.7	&--2.1	&	7.1	&	2330.3	& O, S\\
36b	&	H	&	11\phn 24\phn 	39.8	&--59\phn19\phn42.5	&	41.4	&--231.5	&	60.5	&	1.2	&--6.7	&	0.2	&	6.8	&	2231.7	& O\\
37	&	RC	&	11\phn 24\phn 	39.9	&--59\phn16\phn54.9	&	41.8	&--63.9	&	906.1	&	1.2	&--1.9	&	2.8	&	3.6	&	1160.0	& O\\
38a	&	H, RC	&	11\phn 24\phn 	40.6	&--59\phn19\phn51.7	&	47.2	&--240.7	&--1.6	&	1.4	&--7.0	&	0.0	&	7.1	&	2326.9	& O\\
38b	&	H, RC	&	11\phn 24\phn 	40.6	&--59\phn19\phn51.7	&	47.2	&--240.7	&	506.4	&	1.4	&--7.0	&	1.6	&	7.3	&	2381.4	& O\\
39	&	RC	&	11\phn 24\phn 	41.1	&--59\phn16\phn40.2	&	51.3	&--49.2	&	1158.9	&	1.5	&--1.4	&	3.6	&	4.1	&	1340.9	& O\\
40	&	H, RC	&	11\phn 24\phn 	41.2	&--59\phn19\phn16.7	&	52.2	&--205.7	&--621.2	&	1.5	&--6.0	&--1.9	&	6.5	&	2106.7	& O\\
41a	&	RC	&	11\phn 24\phn 	41.5	&--59\phn19\phn10.1	&	54.5	&--199.1	&--713.4	&	1.6	&--5.8	&--2.2	&	6.4	&	2084.4	& O\\
41b	&	RC	&	11\phn 24\phn 	41.5	&--59\phn19\phn10.1	&	54.5	&--199.1	&	69.0	&	1.6	&--5.8	&	0.2	&	6.0	&	1959.8	& O\\
42	&	H, RC	&	11\phn 24\phn 	41.9	&--59\phn18\phn50.7	&	57.2	&--179.7	&	1041.1	&	1.7	&--5.2	&	3.2	&	6.3	&	2069.7	& O, S\\
43	&	RC	&	11\phn 24\phn 	42.2	&--59\phn18\phn38.0	&	59.6	&--167.0	&	343.0	&	1.7	&--4.9	&	1.1	&	5.3	&	1716.8	& O\\
44a	&	RC	&	11\phn 24\phn 	42.3	&--59\phn16\phn26.0	&	60.5	&--35.0	&--48.1	&	1.8	&--1.0	&--0.1	&	2.0	&	665.2	& O\\
44b	&	RC	&	11\phn 24\phn 	42.3	&--59\phn16\phn26.0	&	60.5	&--35.0	&	1062.5	&	1.8	&--1.0	&	3.3	&	3.8	&	1252.7	& O\\
45a	&	H	&	11\phn 24\phn 	42.7	&--59\phn16\phn01.9	&	63.6	&--10.9	&	482.4	&	1.9	&--0.3	&	1.5	&	2.4	&	779.7	& O, S\\
45b	&	H	&	11\phn 24\phn 	42.7	&--59\phn16\phn01.9	&	63.6	&--10.9	&	1108.8	&	1.9	&--0.3	&	3.4	&	3.9	&	1266.8	& O\\
46a	&	RC	&	11\phn 24\phn 	43.2	&--59\phn16\phn15.1	&	67.6	&--24.1	&	102.3	&	2.0	&--0.7	&	0.3	&	2.1	&	688.6	& O\\
46b	&	RC	&	11\phn 24\phn 	43.2	&--59\phn16\phn15.1	&	67.6	&--24.1	&	939.2	&	2.0	&--0.7	&	2.9	&	3.6	&	1160.1	& O\\
47a	&	RC	&	11\phn 24\phn 	43.2	&--59\phn17\phn46.0	&	67.8	&--115.0	&--53.8	&	2.0	&--3.3	&--0.2	&	3.9	&	1267.6	& O\\
47b	&	RC	&	11\phn 24\phn 	43.2	&--59\phn17\phn46.0	&	67.8	&--115.0	&	680.4	&	2.0	&--3.3	&	2.1	&	4.4	&	1437.7	& O\\
48a	&	H	&	11\phn 24\phn 	43.3	&--59\phn16\phn31.1	&	68.1	&--40.1	&--340.6	&	2.0	&--1.2	&--1.0	&	2.5	&	823.1	& O, S\\
48b	&	H	&	11\phn 24\phn 	43.3	&--59\phn16\phn31.1	&	68.1	&--40.1	&	26.7	&	2.0	&--1.2	&	0.1	&	2.3	&	749.8	& O\\
48c	&	H	&	11\phn 24\phn 	43.3	&--59\phn16\phn31.1	&	68.1	&--40.1	&	986.2	&	2.0	&--1.2	&	3.0	&	3.8	&	1238.6	& O, S\\
49a	&	RC	&	11\phn 24\phn 	43.6	&--59\phn17\phn27.4	&	70.7	&--96.4	&--0.1	&	2.1	&--2.8	&	0.0	&	3.5	&	1134.2	& O\\
49b	&	RC	&	11\phn 24\phn 	43.6	&--59\phn17\phn27.4	&	70.7	&--96.4	&	999.4	&	2.1	&--2.8	&	3.1	&	4.6	&	1511.7	& O\\
50a	&	RC	&	11\phn 24\phn 	44.5	&--59\phn16\phn46.9	&	77.1	&--55.9	&--63.3	&	2.2	&--1.6	&--0.2	&	2.8	&	905.8	& O\\
50b	&	RC	&	11\phn 24\phn 	44.5	&--59\phn16\phn46.9	&	77.1	&--55.9	&	876.0	&	2.2	&--1.6	&	2.7	&	3.9	&	1258.5	& O\\
51a	&	H	&	11\phn 24\phn 	44.6	&--59\phn15\phn36.5	&	78.2	&	14.5	&	38.9	&	2.3	&	0.4	&	0.1	&	2.3	&	755.9	& O, S\\
51b	&	H	&	11\phn 24\phn 	44.6	&--59\phn15\phn36.5	&	78.2	&	14.5	&	727.1	&	2.3	&	0.4	&	2.2	&	3.2	&	1048.1	& O\\
52a	&	RC	&	11\phn 24\phn 	44.7	&--59\phn16\phn34.7	&	79.0	&--43.7	&--162.7	&	2.3	&--1.3	&--0.5	&	2.7	&	872.1	& O\\
52b	&	RC	&	11\phn 24\phn 	44.7	&--59\phn16\phn34.7	&	79.0	&--43.7	&	987.4	&	2.3	&--1.3	&	3.0	&	4.0	&	1307.2	& O\\
53a	&	H, RC	&	11\phn 24\phn 	44.8	&--59\phn16\phn22.7	&	80.1	&--31.7	&--2.0	&	2.3	&--0.9	&	0.0	&	2.5	&	816.8	& O\\
53b	&	H, RC	&	11\phn 24\phn 	44.8	&--59\phn16\phn22.7	&	80.1	&--31.7	&	948.1	&	2.3	&--0.9	&	2.9	&	3.8	&	1251.5	& O\\
54a	&	H	&	11\phn 24\phn 	45.0	&--59\phn16\phn48.9	&	81.1	&--57.9	&--143.8	&	2.4	&--1.7	&--0.4	&	2.9	&	955.9	& O, S\\
54b	&	H	&	11\phn 24\phn 	45.0	&--59\phn16\phn48.9	&	81.1	&--57.9	&	996.1	&	2.4	&--1.7	&	3.1	&	4.2	&	1373.0	& O, S\\
55	&	RC	&	11\phn 24\phn 	45.1	&--59\phn15\phn52.7	&	82.1	&--1.7	&	292.1	&	2.4	&--0.1	&	0.9	&	2.6	&	832.2	& O\\
56a	&	RC	&	11\phn 24\phn 	45.4	&--59\phn16\phn03.3	&	84.0	&--12.3	&--21.1	&	2.4	&--0.4	&--0.1	&	2.5	&	805.4	& O\\
56b	&	RC	&	11\phn 24\phn 	45.4	&--59\phn16\phn03.3	&	84.0	&--12.3	&	674.3	&	2.4	&--0.4	&	2.1	&	3.2	&	1050.2	& O\\
57	&	H, RC	&	11\phn 24\phn 	46.4	&--59\phn15\phn44.4	&	92.0	&	6.6	&	397.7	&	2.7	&	0.2	&	1.2	&	2.9	&	961.4	& O, S\\
58a	&	H	&	11\phn 24\phn 	46.9	&--59\phn17\phn07.8	&	95.9	&--76.8	&--343.6	&	2.8	&--2.2	&--1.1	&	3.7	&	1215.5	& O\\
58b	&	H	&	11\phn 24\phn 	46.9	&--59\phn17\phn07.8	&	95.9	&--76.8	&	1006.8	&	2.8	&--2.2	&	3.1	&	4.7	&	1540.5	& O\\
59	&	H, RC	&	11\phn 24\phn 	47.6	&--59\phn15\phn24.0	&	100.8	&	27.0	&	193.0	&	2.9	&	0.8	&	0.6	&	3.1	&	1008.7	& O, S\\
60	&	H, RC	&	11\phn 24\phn 	48.1	&--59\phn15\phn17.8	&	105.4	&	33.2	&	102.4	&	3.1	&	1.0	&	0.3	&	3.2	&	1053.3	& O, S\\
61a	&	H	&	11\phn 24\phn 	48.6	&--59\phn16\phn21.0	&	108.7	&--30.0	&--368.9	&	3.2	&--0.9	&--1.1	&	3.5	&	1131.5	& O\\
61b	&	H	&	11\phn 24\phn 	48.6	&--59\phn16\phn21.0	&	108.7	&--30.0	&	761.3	&	3.2	&--0.9	&	2.3	&	4.0	&	1312.9	& O\\
62a	&	H	&	11\phn 24\phn 	48.6	&--59\phn15\phn41.4	&	109.2	&	9.6	&	37.8	&	3.2	&	0.3	&	0.1	&	3.2	&	1041.0	& O\\
62b	&	H	&	11\phn 24\phn 	48.6	&--59\phn15\phn41.4	&	109.2	&	9.6	&	417.2	&	3.2	&	0.3	&	1.3	&	3.4	&	1120.8	& O\\
63a	&	H	&	11\phn 24\phn 	48.8	&--59\phn16\phn16.2	&	110.7	&--25.2	&--394.5	&	3.2	&--0.7	&--1.2	&	3.5	&	1146.8	& O\\
63b	&	H	&	11\phn 24\phn 	48.8	&--59\phn16\phn16.2	&	110.7	&--25.2	&	869.6	&	3.2	&--0.7	&	2.7	&	4.2	&	1384.1	& O\\
64	&	H	&	11\phn 24\phn 	49.1	&--59\phn14\phn26.8	&	112.7	&	84.2	&--175.8	&	3.3	&	2.4	&--0.5	&	4.1	&	1346.4	& O, S\\
65	&	RC	&	11\phn 24\phn 	49.1	&--59\phn15\phn05.3	&	112.9	&	45.7	&	15.0	&	3.3	&	1.3	&	0.0	&	3.5	&	1155.5	& O\\
66a	&	H, RC	&	11\phn 24\phn 	51.4	&--59\phn14\phn44.0	&	130.5	&	67.0	&--1072.0	&	3.8	&	1.9	&--3.3	&	5.4	&	1757.1	& O, S\\
66b	&	H	&	11\phn 24\phn 	51.4	&--59\phn14\phn44.0	&	130.5	&	67.0	&--5.0	&	3.8	&	1.9	&	0.0	&	4.3	&	1392.2	& O, S\\
67	&	RC	&	11\phn 24\phn 	51.5	&--59\phn14\phn37.4	&	130.9	&	73.6	&--1056.5	&	3.8	&	2.1	&--3.2	&	5.4	&	1773.5	& O\\
68a	&	H	&	11\phn 24\phn 	51.9	&--59\phn14\phn31.6	&	134.1	&	79.4	&--46.7	&	3.9	&	2.3	&--0.1	&	4.5	&	1479.5	& O\\
68b	&	H, RC	&	11\phn 24\phn 	51.9	&--59\phn14\phn31.6	&	134.1	&	79.4	&--1084.4	&	3.9	&	2.3	&--3.3	&	5.6	&	1833.8   & O\\
\enddata
\tablenotetext{a}{ Labels are in order of increasing RA.  Where two knots have the same RA and Dec (but different radial velocity), the knots are labeled with a letter.}
\tablenotetext{b}{ Fields are labelled as "H" if observed with Hydra, "RC" if observed with RC spectrograph, and "H,RC" if observed with both.}
\tablenotetext{c}{ Displacements are with respect to the center of expansion R.A. (2000.) = 11:24:34.4, decl. (2000.) = -59:15:51 from \citet{winkler09}.}
\tablenotetext{d}{ Physical lengths assume a distance of $d_6=6$ kpc. }
\tablenotetext{e}{ Distance traveled along the line of sight assumes an expansion age of $\tau_3=3000$ yrs. }
\tablenotetext{f}{ Total distance traveled by the knot $R=\left(\Delta x^2+\Delta y^2+\Delta z^2\right)^{0.5}$}
\tablenotetext{g}{ Total velocity of the knot $V=\left(v_x^2+v_y^2+v_r^2\right)^{0.5}$}
\tablenotetext{h}{ O $\implies $ \oii\ lines only; O, S $\implies $\oiii\ and \sii\ lines.}
\end{deluxetable*}


\section{Three-Dimensional Kinematics}
\label{sec:res}
For this paper, we consider a free-expansion model in which ejecta are traveling outward from a common expansion center, found by WTRL to be:
\begin{eqnarray}
\label{eq:ra}
\mathrm{R.A.} (2000.) = 11^h24^m34\fs4, \\
\nonumber \mathrm{Decl.} (2000.) = -59\degr 15\arcmin 51\arcsec.
\end{eqnarray}
The center of the outer radio shell found by \citet{gaensler03} is only 3$\arcsec$ southeast of this optical expansion center.  For each knot we measured the location in the plane of the sky and determined its $x$- and $y$-position relative to the expansion center.  We report the projected positions and the radial velocities for all knots in Table \ref{tab:vrad}.

WTRL determined   (see their Figure 6) that for each knot the distance traveled from a common expansion center in the plane of the sky was proportional to the transverse velocity of that knot.  Similarly, we assume that the knots are not decelerating, and thus the distance traveled along the line of sight is proportional to the radial velocity.  Throughout this analysis, we adopt a distance of 6 kpc and an expansion age of 3000 years (WTRL).  Transverse velocities in the plane of the sky are scaled by distance and time according to: 
\begin{equation}
v_{x,y} \equiv (v_{x,y})_{6,3} \left(\frac{d_6}{\tau_3} \right)
\end{equation}
Distance traveled in the z-direction is calculated as radial velocity multiplied by the time of expansion, $\tau=3000$\ years. 

\begin{figure}[htbp]
\centering
\includegraphics[scale=.7]{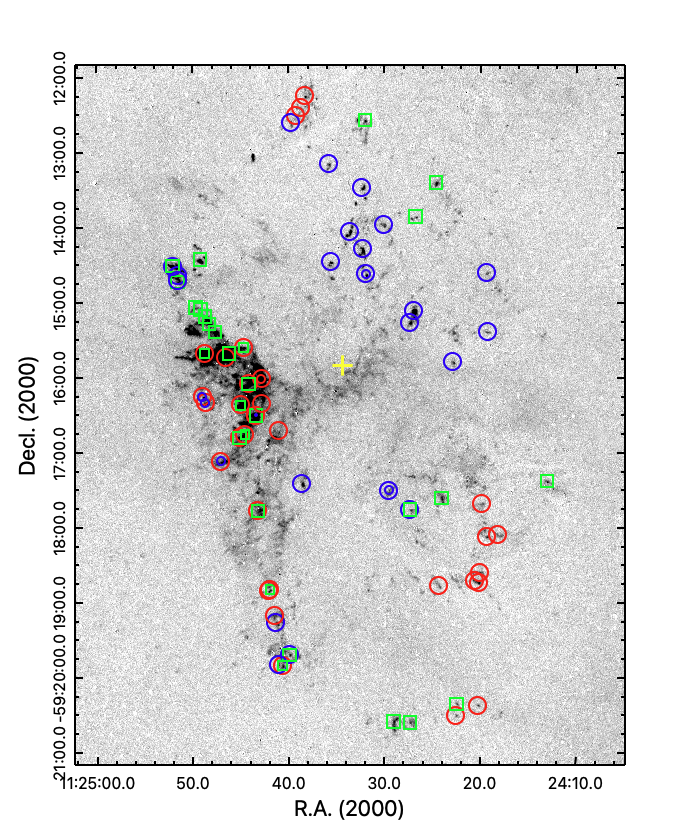}
\caption{Continuum-subtracted \oiii\ image of G292 showing locations of knots for which optical spectra were extracted.  Color coding indicates Doppler shift: red circles indicate knots with $v_{rad} > 330 \kms$, blue circles indicate  $v_{rad} < -330 \kms$,   and green squares indicate $-330 \kms <  $ $v_{rad} $ $< 330 \kms $.  For clarity, smaller symbols are used for some knots with multiple kinematic components.  Knots in the northern region are mostly blue-shifted, while knots in the bright eastern spur and the south are mostly red-shifted. The yellow cross marks the expansion center from WTRL. \label{fig:hrc}}
\end{figure}

\begin{figure}
\centering
\plotone{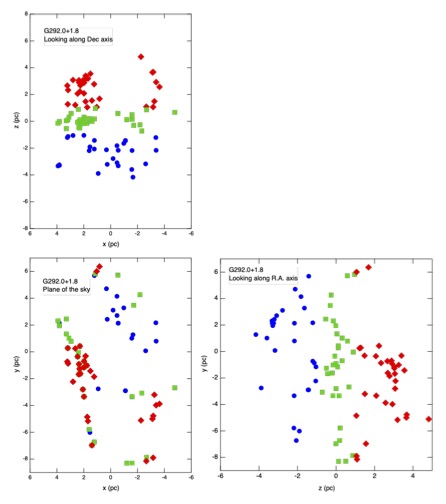}
\caption{The 3-D morphology of optical knots in G292 presented as orthographic projections along the x-, y-, and z-axes.  This assumes a uniform expansion model, with a distance of 6 kpc and an age of 3000 years, as described in the text.  Blue circles represent knots with $v_{rad} < -330 \kms$; green squares, knots with $-330 \kms <$  $v_{rad}$  $< +330 \kms$; red diamonds, knots with $v_{rad} >  330 \kms$. \label{fig:xyz}}
\end{figure}

\begin{figure}
\begin{center}
\includegraphics[width=11cm]{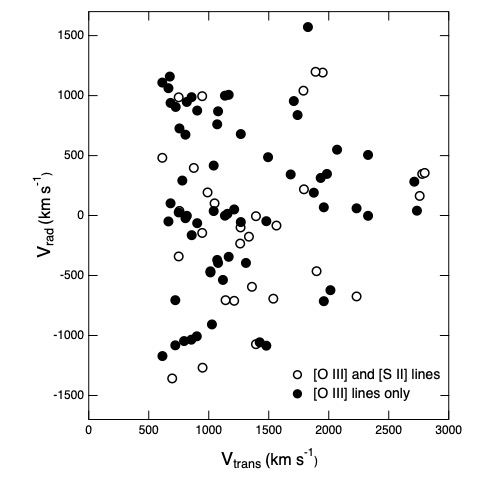}
\end{center}
\caption{Radial velocity $v_{rad}$ versus transverse velocity, for both \oiii-only and \oiii/\sii\ knots.  Transverse velocities are determined as $\sqrt{v_{x}^2+v_{y}^2}$, where velocities have been scaled to a distance of 6.0 kpc and an age of 3000 yr.  For  expansion in which all the knots have close to the same speed, we expect  a spherical shell-like structure, which would appear as an approximately semi-circular distribution, such that the knots with highest radial velocity have the lowest transverse velocity, and vice versa.  Clearly both types of knots in G292 have a wide variety of expansion speeds.}
\label{fig:twodexpand}
\end{figure}

We have thus created a 3-dimensional model of the kinematics of G292 after $\sim 3000$ years of expansion, and we have plotted the $x$-, $y$-, and $z$-coordinates of the ejecta relative to the expansion center, shown in Figure \ref{fig:xyz}.  It is evident that the expansion is not spherical, as seen in Figure \ref{fig:twodexpand}, where for a spherical expansion we would expect that the knots with highest radial velocity to have the lowest transverse velocity, and vice versa.  Given this non-spherical expansion, we therefore inspect the expansion distribution from arbitrary perspectives in order to identify morphological trends.  WTRL showed the two-dimensional projection of G292 in the plane of the sky to be undergoing a roughly bipolar expansion along a North-South axis, with the ejecta spanning $8\farcm4$\ (14.7 pc) North-South versus 5$\arcmin$\ (8.7 pc) East-West.  Incorporating the radial velocities determined spectroscopically,  we find that the ejecta span $\sim 8.7$ pc along the line of sight. We detect mostly blue-shifted ejecta in the northern region, red-shifted ejecta in the dense eastern spur and farthest south, and ejecta relatively stationary along the line of sight projected roughly along an East-West axis.  

\subsection{Reverse shock position and progenitor mass}
Since the optical knots probably represent ejecta that have been excited by the reverse shock, the wide range of knot velocities indicates that the reverse shock has probably penetrated almost to the center of G292.  
\citet{temim22} reached a similar conclusion: that the reverse shock has reached almost all of the PWN (located near the center of the SNR), as well as virtually all of the ejecta, based on their dynamical simulations.  They conclude that the progenitor was of relatively low mass, $M_{ZAMS}\sim 12-16\ M_\sun$ and that most of this mass had been stripped prior to the explosion.  

\added{Whereas our optical data cannot provide additional direct insight regarding the progenitor mass, we note that \citet{narita24} provide additional evidence to constrain the progenitor mass.  Based on XMM observations of the X-ray bright central belt of G292, they found significant nitrogen for the first time and argued that it must originate from shock-heated circumstellar medium produced by the CNO-cycle in the progenitor’s H-burning layer.   They concluded that the progenitor was probably relatively massive, $M_{ZAMS}\gtrsim 30 M_\sun$, and also consistent with a Wolf-Rayet star in a binary system. Most of the mass must have been lost before the explosion, and it has recently been shocked to give the bright central belt of X-ray emission.}


\subsection{An Asymmetrical Explosion}
\begin{figure}
\centering
R\begin{interactive}{animation}{vectors_251216.mp4}
\plotone{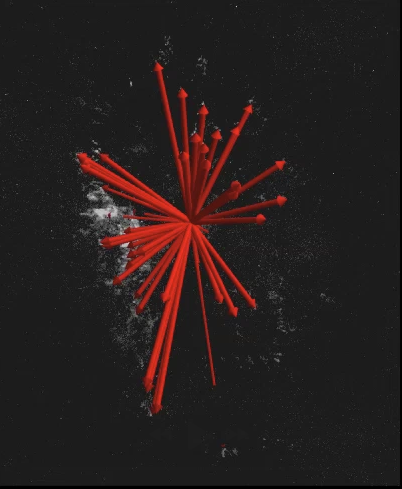}
\end{interactive}
\caption{The 3-D expansion of optical knots in G292 illustrated as vectors that represent the expansion since the explosion. The interactive version of this figure is a 10-sec movie that shows the vectors rotating about a vertical (N-S) axis.
The expanding system of knots has two opposing broad jets, expanding in a roughly N-S direction.\label{fig:vectors}}
\end{figure}

For the ``Crab-like'' remnant 3C~58, \citet{fesen08} proposed a bipolar expansion of high-velocity emission knots ($550 \kms \lesssim v_r \lesssim 1100 \kms $), with a band of high-velocity knots expanding in directions not parallel to the remnant's bipolar radio nebula or to the X-ray emission structures around the central pulsar.  For G292, we find a similar bipolar expansion morphology to that proposed by \citet{fesen08} for high-velocity knots in 3C~58, but with far higher velocities.  As best shown in Fig.~\ref{fig:vectors}, there is a central belt of relatively low velocity knots, perpendicular to the bi-directional jets.  This belt coincides roughly with the bright spur region of G292.

Cas A, the prototypical O-rich SNR, is far younger, brighter, and better studied than G292.  In a detailed kinematic map, \citet{milisavljevic13} found that while most of the ejecta knots and filaments are arranged in ring-like structures, the highest velocity knots are concentrated in  bipolar jets.  But  unlike G292, these knots appear exclusively S-rich (with oxygen weak or absent).  Since they are composed of O-burning products, they probably originated deep within the SN progenitor.


We calculated total 3-dimensional velocities  for the G292 knots and found the velocities to range in magnitude from $670\kms$ to $2800\kms$.  Ranking the knots in order of velocity magnitude, we find a spatial correlation with velocity such that the knots in the dense eastern spur region tend to be traveling slowest, with an average velocity of $1100\kms$. Those in the southwest are the fastest, with an average velocity of $1900\kms$, and those in the north have an average velocity of $1500\kms$.  \citet{ghavamian05} found similar results in their Fabry-Perot measurements,  They also identified the bright eastern-spur as kinematically distinct from other optical filaments, and the velocity distribution we observe seems to coincide with this distinction.


The bi-directional jets in G292 have an axis inclined slightly SE-NW, tilted $\sim 10^{\circ}$ CCW from N-S,  and also tilted $\sim 5^{\circ}$ forward from the plane of the sky.  
We require that the axis for the bipolar jets pass through the center of expansion, but we do not constrain the axis to pass through the pulsar location.  The projection of the ejecta on the equatorial plane gives the most compact arrangement of ejecta, signifying that we are looking down the ``barrels'' of the expansion.  Figs.~\ref{fig:xyz} and \ref{fig:vectors} show this geometry.  As best shown in Fig.~\ref{fig:vectors}, the knots north of the plane are generally blue-shifted, while those south of it are mostly red-shifted.



\subsection{Oxygen-burning Products}
As previously described, we consistently determined positions of knots based on continuum-subtracted \oiii \ images and the Doppler shifts of the strong \oiii \ doublet features for 93 knots.  Of these knots, 32 contained spectra with \sii \ $\lambda\lambda$ \ 6716, 6731 features, reaffirming the presence of sulfur identified by \citet{winkler06} in \sii \ images.  Sulfur is a Si group element that results from O burning, either hydrostatically in the progenitor star or explosively during the SN.  The presence of sulfur concomitant with a lack of \ha \ implies that the optical knots are likely pure ejecta from the oxygen-burning shell of the star.   The presence of sulfur is not, however, correlated with velocity of the ejecta, as we detect sulfur in ejecta with a wide range of velocities, \added {suggesting that  O burning products are mixed in the SN explosion.}

\added{\citet{bhalerao19} have also studied O-burning products in G292, by identifying X-ray knots in {\em Chandra} emission-line images.  They found that virtually all of the \ion{Si}{13}- and \ion{S}{15}-emitting knots are located in the NW hemisphere of G292.  \citet{bhalerao19}  suggest that this is due to an asymmetric explosion, which kicked the pulsar toward the SE, as is observed.  While this seems a reasonable suggestion, the optical data do not provide significant support for it: The majority of the optical \sii-emitting knots are in the northern half of G292, but \sii\ knots are also seen in the south. More importantly, the velocity distribution of all of the optical knots,  illustrated in Figs.~\ref{fig:twodexpand} and \ref{fig:vectors}, does not imply more rapid expansion towards the NW.   }

\section{SUMMARY}
Here we have carried out medium-resolution spectroscopy of 93 optical emission knots in the SNR G292.0+1.8 and have used this to extend our previous proper-motion study    (WTRL) to construct a 3-D picture of the un-decelerated ejecta.
Our optical observations of G292 reveal the kinematics of the O-rich ejecta that are interacting with  the reverse shock of the SN.  Specifically, for 93 knots we detected emission of the strong \oiii \ $\lambda\lambda$ \ 4959, 5007 doublet.  That the knots show a substantial amount of oxygen and lack of hydrogen signifies that these are pure ejecta from the oxygen  shell of the progenitor star, and as pure ejecta are yet uncontaminated by circumstellar and/or interstellar material.  All this is consistent with WTRL, who found that for all knots the velocity is proportional to the distance traveled from the common expansion center, given by equation \ref{eq:ra}.  We summarize our main findings as follows:

1. Ejecta are expanding fastest along a roughly North-South axis, with blue-shifted knots mostly in the northern region and red-shifted knots in the dense eastern spur and in the farthest southern region.  Some of the knots which have traveled farthest south extend beyond the X-ray emitting region shown by \citet{park07}.  Also, when we plot radial velocity versus projected radius, the ejecta are not distributed in a near-spherical shell   as proposed by \citet{ghavamian05}.   The expansion is clearly aspherical, and suggests a bipolar morphology. 

The magnitude of total velocity of the knots also corresponds to the spatial distribution of the projection of the knots on the plane of the sky.  We therefore distinguish between three regions of ejecta: the fastest knots located in the south, fast knots located in the north/northwest, and the slowest knots located in the dense eastern spur.  

2. We find the range of positions along the line of sight (relative to the center of expansion) to be roughly equal to the range of positions in the x-direction in the plane of the sky, indicating that the 
scaling factors for distance ($d_6 \equiv d/(6\ \textrm{kpc}$) and age ($\tau_3 \equiv t/(3000\ \textrm{yr})$) are approximately correct.  Furthermore, when we project the ejecta onto our proposed axis of symmetry, perpendicular to the near N-S bipolar distribution, the ejecta distribution is most compact.

3. Optical emission is the result of dense O-rich ejecta material shocked by the reverse shock, which has penetrated almost the entirety of the SNR.     In addition to the oxygen emission, about a third of the knots we identified also displayed \sii \ $\lambda\lambda$ \ 6716,  6731 features, indicating the presence of S, and likely other Si-group elements,  products of oxygen-burning.  The morphology of the knots we detected is therefore imperative to understanding both the distribution of dense ejecta material from the oxygen-burning shell of the progenitor star, as well as the distribution of  circumstellar material ejected from the progenitor star prior to the SN explosion.  The distribution of the S-containing knots throughout the SNR suggests that the ejecta were well mixed in the SN explosion.
\bigskip

The optimum way to observe the 3-dimensional structure of G292 would be to observe the entire SNR with an integral field unit (IFU) such as the MUSE instrument on the VLT, or a similar instrument.  However, the field for MUSE, even in its "wide field mode," is only $1\arcmin \times 1\arcmin$.  This makes observing the entirety of G292, with a diameter of $\sim 8
\arcmin$ a rather daunting project which has yet to be undertaken.

\begin{acknowledgments}
This research has been sponsored in part by the NSF through grants AST-0307613 and AST-0908566.  Finally, the Middlebury College faculty research fund has partially underwritten the costs of publishing this research. 
\end{acknowledgments}

{\it Facilities:} {NSF NOIRLab's CTIO (4m Blanco, 1.5m)}

\software {SAOimage ds9, IRAF, Igor Pro\textsuperscript{\textregistered}}




\bibliographystyle{aasjournal}

\bibliography{bibmaster}

\end{document}